\documentclass[printer]{aa}  

\usepackage{txfonts}
\usepackage{multicol}
\usepackage{graphicx}
\usepackage{url}

\usepackage{amsmath}  
\usepackage{hyperref}  
\usepackage{psfrag}
\DeclareGraphicsExtensions{.pdf,.png,.jpg,.ps,.eps}
\usepackage{natbib}
\usepackage{subfig}
\usepackage{threeparttable}
\bibpunct{(}{)}{;}{a}{}{,} 

\usepackage[normalem]{ulem}
\usepackage{color}

\begin{document}

\title{Unveiling accretion in the massive YSO G033.3891}  
\subtitle{Spatial and kinematic constraints from the CO bandhead emission}

\author{E.~Koumpia\inst{\ref{inst1}}\fnmsep\thanks{ekoumpia@eso.org}, D.~Sun\inst{\ref{inst3}}, M.~Koutoulaki\inst{\ref{inst2}}, J.~D.~Ilee\inst{\ref{inst2}}, W.-J.~de Wit\inst{\ref{inst1}}, R.~D.~Oudmaijer\inst{\ref{inst2}}, A.~J.~Frost\inst{\ref{inst1}}}

\institute{ESO, Alonso de Córdova 3107 Vitacura, Casilla, 19001, Santiago, Chile\label{inst1}
\and
School of Physics \& Astronomy, University of Leeds, Woodhouse Lane, LS2 9JT, Leeds, UK\label{inst2} 
\and
Physics Department, Princeton University, Princeton, NJ, USA\label{inst3}
}

\date{Received date/Accepted date}

\abstract
{The inner parts of the hot discs surrounding massive young stellar objects (MYSOs) are still barely explored due to observational limitations in terms of angular resolution, scarcity of diagnostic lines and the embedded and rare nature of these targets.}{We present the first K-band spectro-interferometric observations toward the MYSO G033.3891, which based on former kinematic evidence via the CO bandhead emission is known to host an accreting disc.}{Using the high spectral resolution mode (R$\sim$4000) of the GRAVITY/VLTI, we spatially resolve the emission of the inner dusty disc and the crucial gaseous interface between the star and the dusty disc. Using detailed modelling on the K-band dust continuum and tracers known to be associated with the ionised and molecular gaseous interface (Br$\gamma$, CO), we report on the smallest scales of accretion/ejection.}{The new observations in combination with our geometric and kinematic models employed to fit former high spectral resolution observations on the source (R$\sim$30,000; CRIRES/VLTI) allow us to constrain the size of the inner gaseous disc both spatially and kinematically via the CO overtone emission at only 2 au. Our models reveal that both Br$\gamma$ and CO emissions are located well within the dust sublimation radius (5~au) as traced by the hot 2.2~$\mu$m dust continuum.}{Our paper provides the first case study where the tiniest scales of gaseous accretion around the MYSO G033.3891 are probed both kinematically and spatially via the CO bandhead emission. This analysis of G033.3891 stands as only the second instance of such investigation within MYSOs, underscoring the gradual accumulation of knowledge regarding how massive young stars gain their mass, while further solidifying the disc nature of accretion at the smallest scales of MYSOs.}  


\titlerunning{Unveiling accretion in the massive YSO G033.3891} 
\authorrunning{Koumpia et al.} 
 \maketitle

\vspace{0.2cm}

\section{Introduction}

Massive stars ($M>8M_{\sun}$) are among the most influential objects of a galaxy. They exert significant influence over the motion of surrounding material but also shape a galaxy's chemical composition by generating heavier elements. However, despite their importance, their formation and evolution are still not well understood. For objects so large, sustaining mass accretion in the face of substantial radiation pressure poses a challenge.
 
There appears to be an agreement that massive stars can form following the accretion-ejection paradigm \citep[e.g.,][]{Rosen2019,Klassen2016,Hosokawa2010}. The short lifetimes of discs surrounding massive young stars \citep[$<$ 10$^{5}$ yr; e.g.,][]{Kuiper2018} seem to be coupled with the strong photoevaporation wind driving most of the disc dispersal \citep[e.g.,][]{Owen2011,Ercolano2015}.   

\citet{Ilee2013} performed kinematic modelling of the CO bandhead emission towards 20 massive young stellar objects (MYSOs) and found evidence of gaseous discs that 'live' at distances as close as few (sub-)au from the central protostars \citep[see also,][]{Bik2004,Chandler1993,Chandler1995}. Direct spatial information of such small-scale hot discs has been limited to a couple of studies of individual objects via hot dust emission, or the CO bandhead \citep{Kraus2010,Caratti2020}. \cite{Koumpia2021} add a sample of six MYSOs to this dedicated effort. More recently, the Na~I doublet and Helium emission have been found to also originate from such small scales in the interface between the dusty disc and the protostellar objects \citep{Koumpia2023}. 

The CO first overtone (or 'bandhead') emission is not commonly observed in the spectra of MYSOs at low to medium resolutions. Instead, it is detected in only about 25\% of their spectra \citep{Ilee2018}.
Until now, there has not been a study combining kinematic modelling of the molecular gaseous component (CO) using high spectral resolution observations (R$\sim$30,000) with direct spatial information. The only case that used a similar approach was limited to a spectral resolution of R $\sim$4,000 \citep{Caratti2020}. Yet, it is crucial to provide the ultimate proof that not only small-scale gaseous discs surrounding MYSOs exist but they are the pathways to active accretion. 

In addition, observationally all MYSOs appear to show Br$\gamma$ emission in their spectrum \citep{Bunn1995,Frost2021}. The origin of the Br$\gamma$ emission has been previously attributed to 'disc-wind' \citep{Drew1998}, or jet \citep{Caratti2016}. Still, it is not always straightforward to tell those apart, especially when the spectral resolution is not adequate to do so \citep{Koumpia2021}. Given the influence of both mechanisms on the final mass of the central object, the combination of high spectral and spatial resolution is crucial. Interferometry and the unique information provided by the differential phase of GRAVITY in HR mode are powerful tools in telling those processes apart by enabling astrometric solutions in such a high angular resolution.

In this paper, we focus on the MYSO G033.3891 (IRAS 18490+0026), which with a lower-mass limit of M $>$ 12~$M_{\odot}$ and luminosity of L$\sim$1.3$\times$10$^{4}$~$L_{\odot}$ is a good example of this class of objects. G033.3891 is one of the very few MYSOs known to host an accreting gaseous disc as traced via the CO bandhead emission \citep{Wheelwright2010} and is bright enough for high angular resolution observations in the K-band ($K=7^{m}.2$) with GRAVITY on the VLTI (Very Large Telescope Interferometer). \citet{Ilee2013} modelled the observed CRIRES (Cryogenic high-resolution Infrared Echelle spectrograph) spectrum of the source around the bandheads (R$\sim$30,000) and found that the best solution was achieved for a disc with an inner radius of only $\sim$2~au. The inclination of the disc was determined to be approximately $\sim$40$^{\circ}$. 

This work presents the first spatially resolved observations of the hot dust as traced with 2.2~$\mu$m continuum, and the ionised (Br$\gamma$) and molecular (CO) gas content on the source using near-IR interferometry. Our observations allow us to apply detailed geometrical models and to constrain the origin of the Br$\gamma$ emission spatially using astrometry. We finally constrain the origin of the CO bandhead emission towards G033.3891 both spatially and kinematically for the first time.



\section{Methods and observations}


\subsection{GRAVITY observations and data reduction}
\label{obs} 

\begin{figure}
    \centering
    \includegraphics[width=0.5\textwidth]{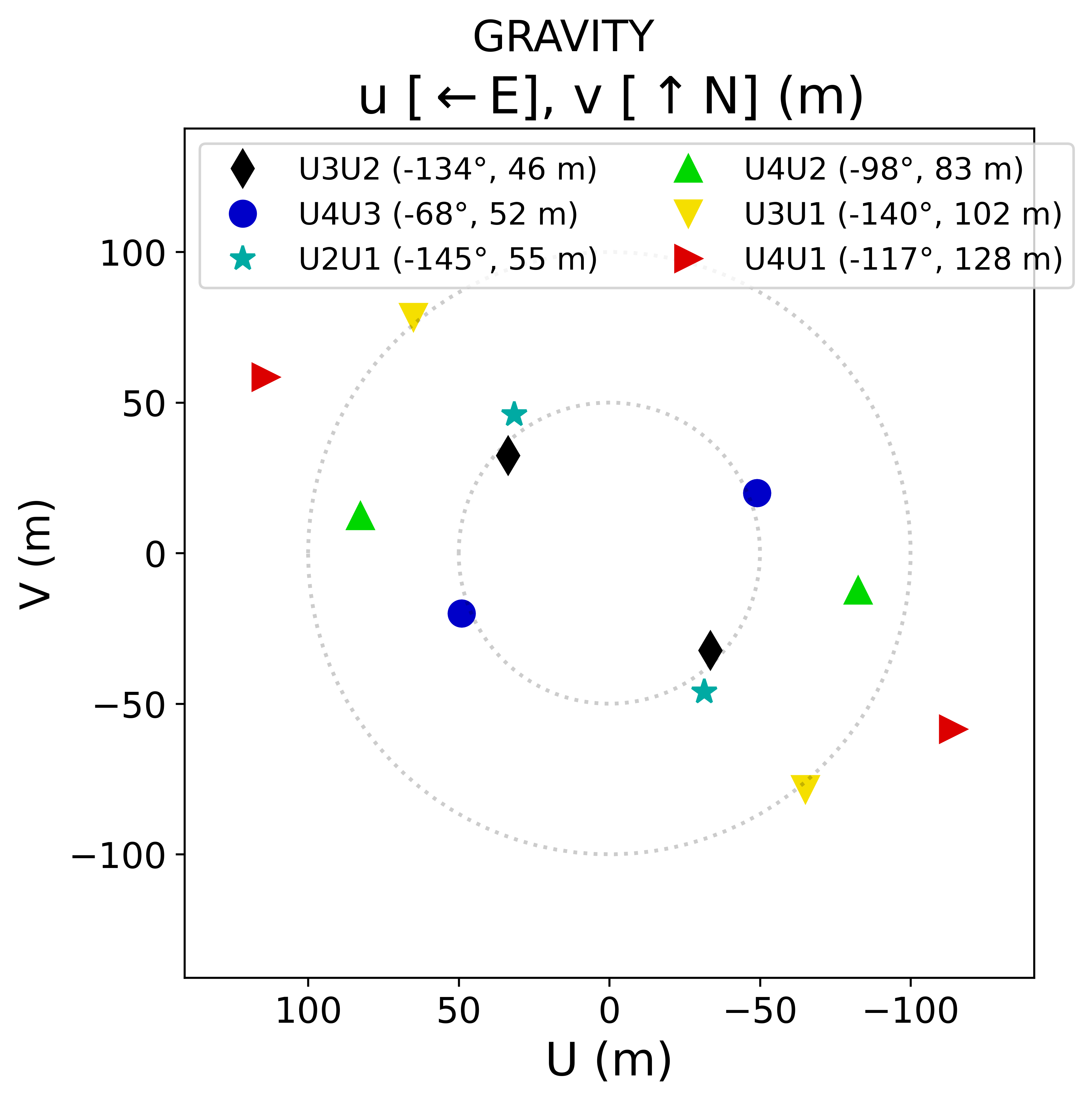}
    \caption{uv-coverage of G033.3891 obtained using GRAVITY/VLTI on UTs. The position angle (in degrees) and projected baseline length (in metres) of each baseline pair are indicated in the legend.}
    \label{uv}
\end{figure}

MSX6C G033.3891+00.1989 (IRAS 18490+0026; RA = 18$^{h}$51$^{m}$33.8$^{s}$, Dec = +00$^{\circ}$29\arcmin51\arcsec.1 [J2000]; Table~\ref{G033_table}) was observed on the 30th of May 2021 with the GRAVITY interferometric instrument \citep{Gravity_Coll2017,Eisenhauer2011} on the VLTI using the four 8.2-m Unit Telescopes (UTs). GRAVITY covers the K-band (1.99~$\mu$m - 2.45~$\mu$m). The deeply embedded nature of the source and the absence of suitable guiding stars on-axis required the use of the Coudé CIAO system for natural star guiding off-axis. The observations were taken using the high spectral resolution in combined polarisation mode (HR; R$\sim$4000, $\Delta$v$\sim$75~kms$^{-1}$). The observing run\footnote{DIT: 30~sec; NDIT (on science): 12; Total Exposure Time of the SCI-CAL sequence: 36 minutes} was executed under good atmospheric conditions with a seeing varying between 0.6\arcsec and 0.8\arcsec and $\tau_{\rm cor}\sim5$~ms. The obtained uv-coverage is presented in Figure~\ref{uv}. The fringes didn't show consistent stable behaviour throughout the entire run, yet a high fringe tracking ratio of 81-97\% could still be obtained.  

HD 174606 (RA = 18$^{h}$51$^{m}$16.67$^{s}$, Dec = +00$^{\circ}$49\arcmin59\arcsec.4 [J2000]) was observed as a standard calibrator under good conditions similar to the science object and showed stable fringes with a fringe tracking ratio of 100\%. HD~174606 is characterised by a K-band magnitude of 7$^{m}$, a spectral type of F3III, and a size of a uniform disc in the K-band of 0.157~mas \citep[i.e., spatially unresolved, JMMC SearchCal;][]{Bonneau2011}.

The observations were reduced and calibrated using the standard GRAVITY pipeline recipes as provided by ESO (version 1.5.4). The instrumental transmission in the spectrum was corrected via the pipeline. The spectra were also corrected for tellurics using the HITRAN models \citep{Rothman2009} via PMOIRED \footnote{PMOIRED is a Python3 module which can be downloaded at https://github.com/amerand/PMOIRED} \citep[similar to Molecfit,][]{Smette2015,Merand2022}. To address continuum irregularities and normalise the spectrum we applied up to a third order polynomial fitting. 

\begin{table*}[ht!]
\caption{The coordinates, bolometric luminosity, K-band magnitude, and distance of G033.3891 as taken from the RMS database. The mass and effective temperature of the source are also listed.}
\resizebox{\textwidth}{!}{
\begin{tabular}{clccccccc}
\hline
Source & R.A. & Dec. & L$_{\rm *}$ & Mass & K-band & Distance & T$_{eff}$ & Simbad name \\
 & (J2000) & (J2000) & (L$_{\sun}$) & (M$_{\sun}$) & (mag) & (kpc) & (K) \\
\hline
G033.3891 & 18:51:33.8 & +00:29:51.1 & 13,000 & 12.0 & 7.2 & 5.0 & 26,000 & MSX6C G033.3891+00.1989  \\ 
\hline
\end{tabular}
}
\tiny Notes: The effective temperature and mass of the source are extrapolated following the methods adopted for main sequence objects \citep{Martins2005}. The mass derived with this approach comes with a high uncertainty of 35 to 50\%. The typical uncertainty on the reported luminosities in RMS is 34\% \citet{Mottram2011}. For massive sources, it has been observed that stellar luminosity accounts for the majority of the total luminosity, as demonstrated by Herbig Ae/Be stars \citep[e.g.][]{Fairlamb2015}. Therefore, the bolometric luminosity can be effectively represented by the stellar luminosity for MYSOs.    
\label{G033_table}
\end{table*}

\subsection{Interferometric observables}

\begin{figure}
    \centering
    \includegraphics[width=\columnwidth]{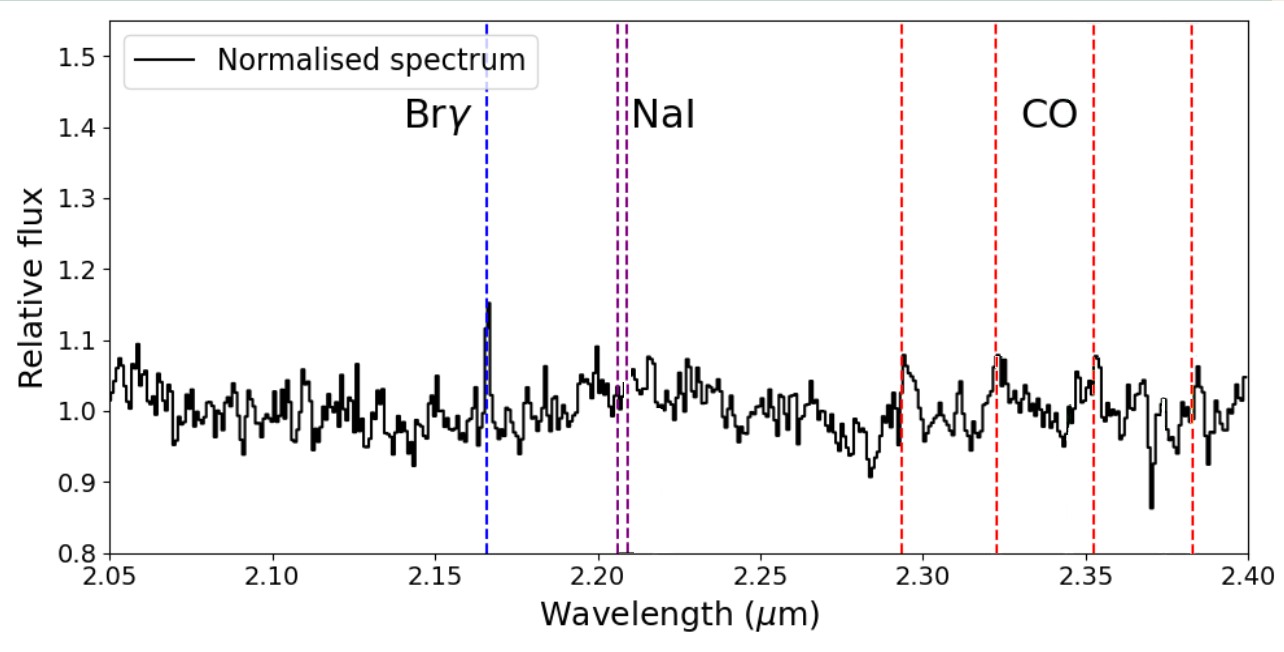}
    \caption{Normalised spectrum for G033.3891 as observed with GRAVITY. The telluric corrected spectrum shows the Br$\gamma$ and the CO bandheads in emission.} 
    \label{spectrum}
\end{figure}

The interferometric observables towards G033.3891 delivered by GRAVITY consist of spectral information, calibrated visibilities, and differential and closure phases in the K-band (2.0-2.4~$\mu$m). The K-band spectrum (Figure~\ref{spectrum}) shows a prominent Br$\gamma$ emission line at 2.167~$\mu$m and the CO bandhead emission around 2.3-2.4~$\mu$m. Based on the presence of Na I in the low-resolution spectra of the source \citep{Pomohaci2017}, the GRAVITY spectra were also inspected for the presence of the Na I doublet emission but without a clear detection. The root-mean-square of the spectrum across the full spectral window is 0.034. An inspection of the calibrated visibilities and phases around the spectral lines of interest, allows us to extract some qualitative information regarding the relative size and symmetric or asymmetric nature of the line emitting regions (Br$\gamma$, CO) compared to the 2.2~$\mu$m continuum emission.

The absolute visibility, also referred to as visibility amplitude V(f,$\lambda$), at the peak intensity of the Br$\gamma$ line shows an increase at all 6 observed baselines compared to the visibilities of the surrounding continuum (Figure~\ref{G033}). At the longer three baselines ($\sim$90~m-130~m) this increase is $>$5\% and reaches up to $\sim$13\% at the longest one (U4U1; 127~m). This is higher than the typical uncertainties introduced by calibrations and the transfer function. The observed increase for the other three baselines is comparable to the typical uncertainties (2-5\%; excluding residual calibration effects). The peak of the line emission coincides with the velocity location corresponding to the peak of the visibility, suggesting that the observed variations at all baselines are due to a real geometrical effect. Larger visibilities along emission lines are suggestive of a smaller emitting area compared to that of the dust continuum. Therefore, the Br$\gamma$ emitting region is more compact compared to the 2.2~$\mu$m continuum.

The differential phases as obtained with GRAVITY show variations around the Br$\gamma$ emission up to 2.8$^{\circ}$. The observed changes suggest an offset between the photocentre of the 2.2$\mu$m continuum emission and that of the Br$\gamma$ emitting region. On the other hand, no variations are observed in the observed closure phases around the line emission, suggesting an axisymmetric nature of the ionised gas emitting region.  

Examining the calibrated visibilities around the CO bandhead emission, there is an increase compared to the surrounding continuum at 2 baselines (U4U1, U4U2), which is $\sim$ 5-7\% at the longest baseline. The consistency in the location of the peak visibility with that of the emitting components of the CO bandheads demonstrates that the observed variation is real. 

The absolute visibilities indicate that the CO bandhead arises from a more compact area compared to the dust continuum emission. In Sec.~\ref{sec:pmoired}, we employ detailed geometric modelling to constrain the sizes of the continuum, Br$\gamma$ and CO bandhead emission quantitatively. 

There are no significant variations in the phases (closure or differential) around the CO bandhead emission compared to the continuum, suggesting that the molecular gas component shares the same photocentre as that of the continuum and the emitting region is symmetric.   
This is only the third case of an MYSO known in literature with spatially resolved observations of the CO bandhead emission \citep[NGC~2024~IRS~2, IRAS~13481-6124;][]{Caratti2020,Koumpia2023}.

\section{Results}

\subsection{Geometric modelling}

\label{sec:pmoired}

We employ geometric modelling of the GRAVITY observations of G033.3891 to characterise the size and morphology of the hot dust emission, as traced by the 2.2~$\mu$m continuum, the Br$\gamma$, and the CO bandhead emission, all captured within the K-band spectrum. 

The present analysis utilises the specialised tool PMOIRED (see also, Sect.~\ref{obs}), designed for parametric polychromatic modelling of spectro-interferometric data. The code enables the simulation of interferometric observables for diverse models of standard brightness distribution blocks (e.g., Gaussian, disc, ring) and any combination thereof. The PMOIRED model composition assumes that the three-dimensional flux can be expressed as
\begin{equation}
    F(x, y, \lambda) = I(x, y) \times S(\lambda)
\end{equation}
where \(x\) and \(y\) represent the angular coordinates on the sky, and \(\lambda\) is the wavelength. Here, \(I\) represents the achromatic (model-dependent; wavelength-independent) spatial distribution of the component, while \(S\) represents its spectrum (flat or smooth function for the continuum emission; wavelength and line width - dependent for the line emission). This formulation allows for the separate consideration of spatial structure (\(I\)) and spectral characteristics (\(S\)) when modelling astrophysical objects \citep[see also,][]{Monnier2003}.

Subsequently, the user can determine the model that best aligns with the observations by evaluating the optimal fit through a least-square approximation and a bootstrapping estimation of the uncertainties, as retrieved from the covariance matrix. The correlation matrix is crucial for ensuring that the parameters are independent of each other. A high correlation (e.g., greater than 90\%) suggests that the parameters are degenerate, either due to an inadequate model parametrisation or because the data cannot distinguish between the two parameters. Our best-fit results as presented below, do not suffer from degenerate solutions.

Our visibility models follow the standard approach for modelling YSOs, including a point source representing the unresolved star in the 2.2~$\mu$m emission together with an extended emission component \citep[e.g.,][]{Kraus2008,Tatulli2008}. This approach ensures a comprehensive representation of the circumstellar continuum emission around an MYSO.
 
\subsubsection{K-band continuum}
\label{geo_cont}

We first constrain the size and structure of the K-band emission. Our fitting process encompasses various combinations of brightness distributions, including Gaussian, disc, and ring profiles (see Table~\ref{mod_par}). To limit the degree of complexity during geometric modelling, in the case of a ring or a disc, a face-on orientation is adopted. 

The optimal fit to the 2.2~$\mu$m continuum emission is characterised by a reduced $\chi^2$ value of 1.2, and it is achieved by adopting a ring of an inner diameter of 1.95$\pm$0.08~mas and an outer diameter of 2.8$\pm$0.06~mas. The reported size represents the best-fit result obtained from fitting the entire K-band continuum, assuming that its size and geometry exhibit minimal variation across the observed spectral range. 
At a source distance of 5~kpc \citep[determined kinematically in RMS survey;][]{Lumsden2013}, the inner and outer radius of the ring are determined to be 5~au and 7~au respectively. The inner ring radius is in perfect alignment with the predicted dust sublimation radius (R$_{dust}$) of the source, which is computed to be 5~au (or else 1~mas at 5~kpc) for a source $\sim$1.3$\times$10$^{4}$~L$_\odot$, and a dust sublimation temperature of 1300~K \citep[e.g. olivines, silicates; see][]{Isella2005,Kama2009,McClure2013}. The dust sublimation radius (R$_{dust}$) is estimated from the dust sublimation temperature, T$_{sub}$, and the stellar luminosity, L$_{*}$, using:

\begin{equation}
R_{dust} = \sqrt{\frac{L*}{16 \pi \sigma_{SB} T_{sub}^4}}
\end{equation}

\hspace{-0.5cm}where $\sigma_{SB}$ is the Stefan-Boltzmann constant.


The simultaneous fit of the visibilities and the closure phases (-2.5$^{\circ}$ $<$ CP $<$ 3$^{\circ}$; typical uncertainties $<$ 1.5$^{\circ}$) of the continuum emission required the introduction of an asymmetric emission. In PMOIRED it is possible to simulate deformed or asymmetric components analytically by the introduction of a 'slant' in a specific direction. The geometric model that best fits the interferometric observables of the continuum required the inclusion of a slant of 1.2 at a position angle of $\sim$70$\pm$6$^\circ$ (Fig~\ref{G033}).  

We conclude that the 2.2~$\mu$m emission traces the annulus corresponding to the inner part of the hot dusty disc surrounding this MYSO. This conclusion is based 
on the temperatures of material that the 2.2~$\mu$m emission is sensitive to ($\sim$1300~K), the ring-shaped morphology of its emitting region derived from our geometrical models, and the observation that the determined inner radius is consistent with the predicted dust sublimation radius of this system.  

\subsubsection{Br$\gamma$ emission}
\label{geo_br}

We employ a similar approach for constraining the brightness distribution of the Br$\gamma$ emission as for the continuum emission. Specifically, we explored diverse combinations of brightness distributions (Gaussian, disc, ring). To limit the number of free parameters during the fitting process of both the line and continuum emissions, we constrained the size and geometry of the continuum emission to be identical to the best-fit result for 2.2~$\mu$m described in Sect.~\ref{geo_cont}.

To derive the size of the line emitting region the actual visibilities of the line need to be extracted by subtracting the continuum contributions. This is done by using Eq.~\ref{eq2}, which describes the total visibility (V$_{line+cont}$) in terms of continuum and line visibilities (V) and fluxes (F) for multi-component sources, solving for (V$_{line}$) as follows:

\begin{equation}
V_{\text{line+cont}} = \frac{V_{\text{cont}} \times F_{\text{cont}} + V_{\text{line}} \times F_{\text{line}}}{F_{\text{cont}} + F_{\text{line}}}
\label{eq2}
\end{equation}

The best-fit for the Br$\gamma$ emission corresponds to a Gaussian brightness distribution (with a reduced $\chi^2$ of approximately 3.2), resulting in a FHWM of 0.66$\pm$0.06 mas (equivalent to 3.3 au). To accurately reproduce the observed changes in differential phases along the Br$\gamma$ emission, we allow the central position of the Gaussian distribution to be a free parameter, rather than fixing it to the same location as the photocentre of the continuum. Introducing an offset between the photocentre of the Br$\gamma$ emission and that of the continuum is necessary to accurately reproduce the observed variations in differential phases, as shown in the top panels of Figure~\ref{G033}. 

The best-fit results in an offset between the photocentre of the Br$\gamma$ emission and that of the compact continuum of about 0.4$\pm$0.1 mas (or 2 au) towards the North-East at PA$\sim$41.4$\pm$3.3$^\circ$. The lack of closure phase variations along the Br$\gamma$ emission suggests that there is no necessity to introduce an asymmetry in its brightness distribution.

The geometrical models demonstrate that the Br$\gamma$ emission comes from an area which is significantly more compact compared to the continuum, well within the inner radius of the hot dusty disc. The observed offset between the Br$\gamma$ photocentre and that of the continuum is found in the same direction as the brighter part of the disc and is indicative of some degree of diversion from a face-on geometry.

\subsubsection{Molecular gas emission}

\label{co_geo}

To model the emission of the CO bandheads (2.29~$\mu$m, 2.32~$\mu$m, 2.35~$\mu$m, 2.38~$\mu$m), we follow a similar approach to that used for Br$\gamma$ and the continuum. To minimise the number of free parameters during the fitting we adopted the size and geometry of the continuum emission to match the best-fit result obtained for the 2.2~$\mu$m emission described in Sect.~\ref{geo_cont}. We then tested the three standard brightness distribution models (Gaussian, uniform disc, ring) to fit the spectro-interferometric observables around the CO emission lines.

The best-fit model (reduced $\chi^2 \sim$ 3.0) is achieved for a ring brightness distribution with an inner diameter of 1.10$\pm$0.12~mas (5.5$\pm$0.6~au), or else an inner radius of 0.55$\pm$0.06~mas ($\sim$2.8$\pm$0.3~au). The preference for the ring model is also justified based on physical expectations. CO bandheads are generally understood to originate from a circumstellar disc with an inner cavity \citep[e.g.,][]{Carr1993,Tatulli2008,Ilee2013}, naturally forming a ring-like structure. 

Constraining the outer radius of the ring appeared to be the most difficult parameter to implement in the fitting process. When considering the 'thickness' of the ring (in PMOIRED this reflects the radial extension of the ring) as a free parameter, convergence of the code led to degenerate solutions, oscillating between an infinitely thin (model parameter 'thick' $<$ 0) or infinitely thick ('thick' $>$ 1) ring. To address this issue and proceed with the fitting process, we ran multiple fits by constraining the thickness of the ring to multiples of 0.1 (up to 1) at a time. Upon fixing the thickness at 0.2 (corresponding to an outer radius of 3.4~au), the best-fit was achieved. By finally fixing the thickness of the ring at this value we could further facilitate a robust fit to all observables leaving the rest of the parameters free. This approach results in a final geometry that does not violate the mathematical solution of a physical system. We note that the goodness of the fit is not as strong for three baselines (U3U2, U4U3 and U2U1) compared to the others. These baselines correspond to shorter baseline lengths (46-55~m), which trace the larger spatial scales, and it likely reflects the difficulty encountered during the fitting process in accurately modelling the outer radius of the ring.

No significant changes in the differential or closure phases along the CO bandheads are observed, implying that the CO emission shares the same photocentre as the continuum and in addition, there is no need to implement an asymmetry on the flux distribution of the emitting area. 

\begin{figure*}
\begin{center}
\includegraphics[scale=0.45]{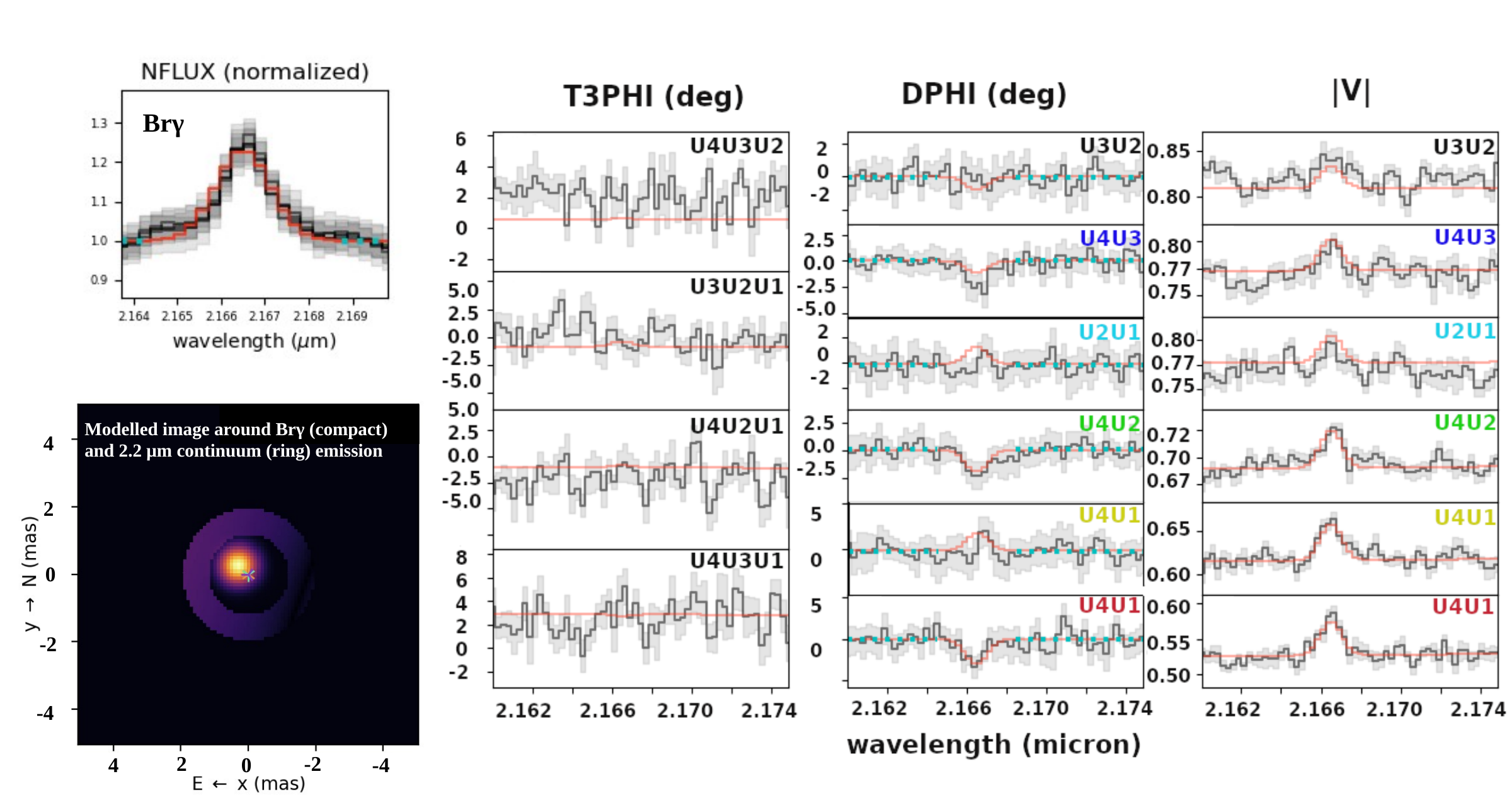} \\
\includegraphics[scale=0.45]{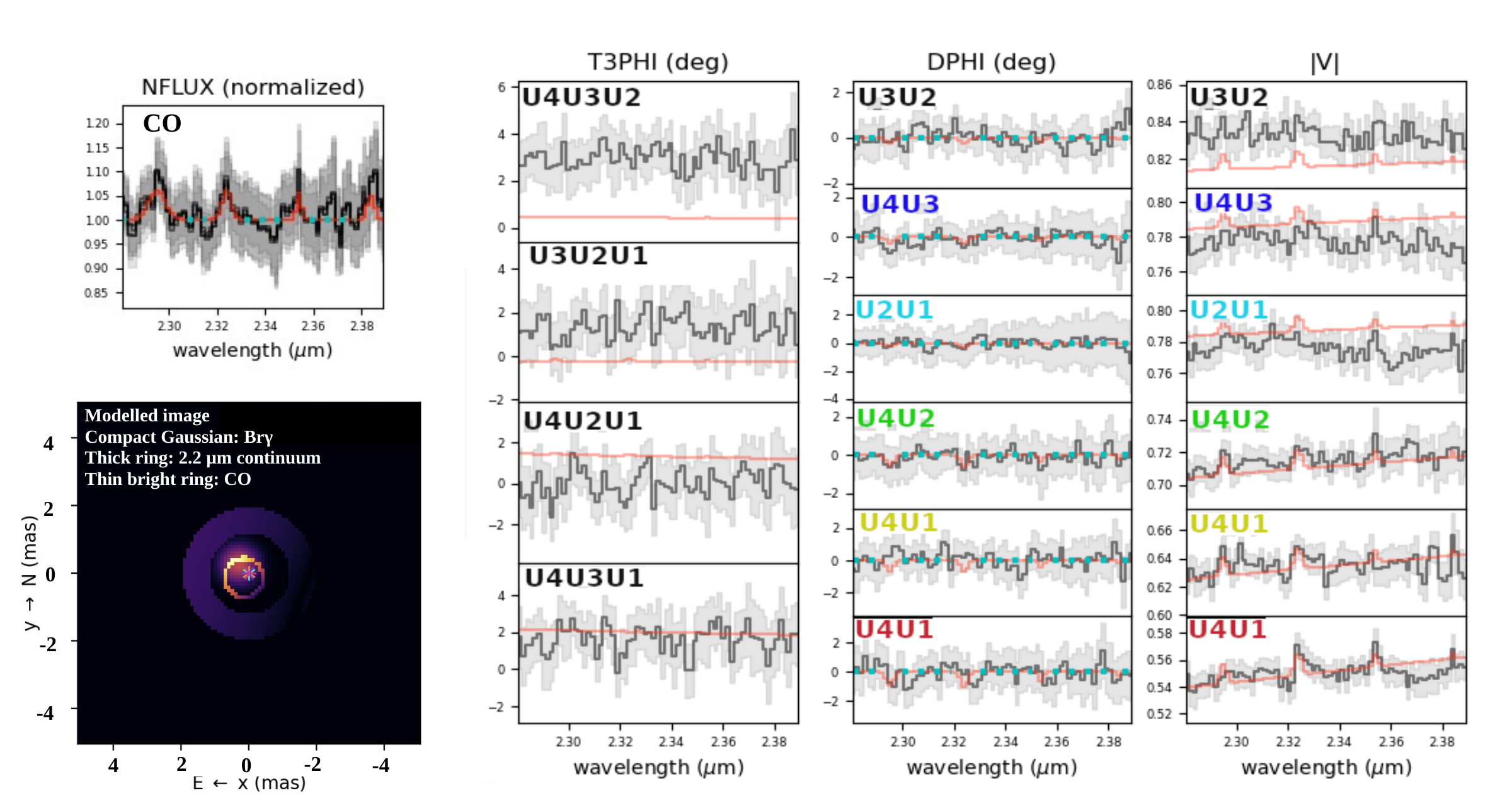}
\end{center}
\caption{Model image of G033.3891 of the 2.2~$\mu$m continuum emission (thick ring) combined with the Br$\gamma$ (compact Gaussian) and CO bandhead molecular emission (thin bright ring). The modelled image of the CO emission is the average of the individual images of the CO bandheads extracted at their central wavelengths (2.29~$\mu$m, 2.32~$\mu$m, 2.35~$\mu$m, 2.38~$\mu$m). The total geometric model (red line) fits well all interferometric observables (closure phases, T3PHI; differential phases, DPHI; visibilities, V; normalised flux, NFLUX) of the continuum, Br$\gamma$ and CO bandheads. PMOIRED normalises the flux from both the data and the model using the same algorithm. The grey shades represent the systematic errors of the interferometric observables, taking into account the calibration and transfer function uncertainties. The colour scheme of the different baselines follows Figure~\ref{uv}.}
\label{G033}
\end{figure*}

The final geometrical model is shown in Figure~\ref{G033}, and it combines the continuum, the Br$\gamma$ and CO bandhead emission (see also Table~\ref{mod_par}). We conclude that the CO emission is similar in size to the Br$\gamma$ emitting region, both significantly more compact compared to the dust sublimation radius, therefore tracing the star/disc interface. 
While our models have successfully constrained the inner radius of the CO ring structure, the robust determination of the molecular content's outer extent remains challenging due to instrumental limitations, such as the spatial resolution, the limited interferometric field of view and the wavelength coverage of GRAVITY.

\begin{table*}[t]
\centering
\caption{Parameters of the best fit to the interferometric observables of the 2.2$\mu$m, Br$\gamma$ and CO emission towards G033.3891.}
\begin{threeparttable}
\begin{small}
\begin{tabular}{lcccccccc}
\hline
  & Model & Size  & PA (slant) & Offset$^{[b]}$ & PA (offset)t & Flux w. & Flux w. & Red. $\chi^{2}$$^{[e]}$ \\
& & (mas) & (degrees) & (mas) & (degrees) & (point) & (extended) &  \\
\hline
2.2~$\mu$m  & Ring$^{[c]}$ & R$_{in}$ - R$_{out}$: 1.00$\pm$0.04 - 1.40$\pm$0.03  & 70$\pm$6 & 0 & na$^{[a]}$ & 0.5$\pm$0.1 & 0.36$\pm$0.04$^{[f]}$ & 1.2 \\
2.2~$\mu$m  & Gaussian & FWHM: 0.95$\pm$0.07 & na & 0 & na  & na & 1.0 $^{[f]}$ & 2.2 \\
2.2~$\mu$m  & Uniform disc & Radius: 1.09$\pm$0.01 & na & 0 & na  & 0.67$\pm$0.04 & 0.17$\pm$0.01$^{[f]}$ & 1.8 \\
Br$\gamma$  & Ring & R$_{in}$ - R$_{out}$: 1.91$\pm$0.4 - 2.39$\pm$0.6  & na & 0.61$\pm$0.07 & 36$\pm$6  & na & 1.0 & 3.5 \\
Br$\gamma$ & Gaussian$^{[d]}$ & FWHM: 0.66$\pm$0.06   &  na  & 0.4$\pm$0.1 & 41.4$\pm$3.3  & na & 1.0 & 3.2 \\
Br$\gamma$ & Uniform disc & Radius: 1.52$\pm$0.08 & na  & 0.26$\pm$0.03 & 67.4$\pm$6.5  & na & 1.0 & 4.0 \\
CO  &  Ring  &  R$_{in}$ - R$_{out}$: 0.55$\pm$0.06 - 0.75 (fixed) & 70$\pm$6 & 0 & na  & na & 1.0 & 3.0 \\
CO  &  Gaussian  & FWHM: 0.79$\pm$0.02 & na & 0 & na & na & 1.0 & 4.6  \\
CO  &  Uniform disc  & Radius: 0.65$\pm$0.04 & na & 0 & na  & na & 1.0 & 4.5 \\
\hline
\end{tabular}
\begin{tablenotes}
\item [.] [a] na: not applicable. [b] The offset is measured with respect to the continuum photocentre. [c] For the entire analysis we adopt the solution of the ring brightness distribution for the 2.2~$\mu$m continuum and the CO bandheads emission. [d] For the entire analysis we adopt the solution of the Gaussian distribution for the Br$\gamma$ emission. [e] The reported reduced $\chi^{2}$ values correspond to the total fit of all interferometric observables combined. [f] An additional, over-resolved (background) component, accounting for 16\% of the total flux contribution, was applied to all geometric models of the continuum.
\end{tablenotes}
\end{small}
\end{threeparttable}
\label{mod_par}
\end{table*}

\subsection{Kinematic modelling of the CO bandhead}

\begin{figure}
    \centering
    \includegraphics[width=0.5\textwidth]{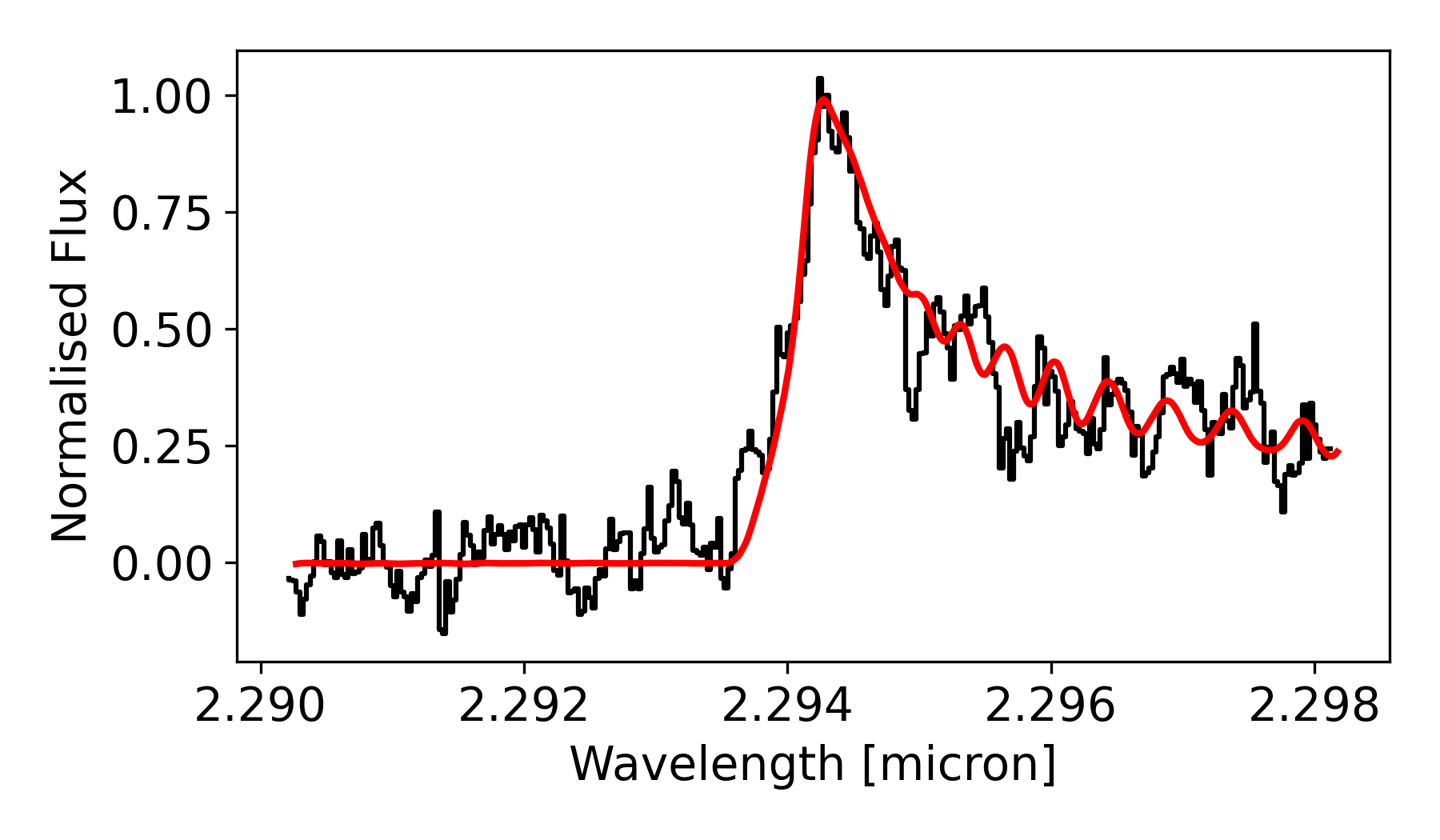}
    \caption{VLT/CRIRES observation of the CO bandhead in black ($R\sim30$,000) overlaid with the best fitting disc model in red \citep[adapted from][]{Ilee2013}.}
    \label{G033-CRIRES}
\end{figure}

In addition to the direct spatial information we retrieved from the GRAVITY dataset in Sect.~\ref{co_geo} regarding the CO size and distribution, we obtain an independent measurement via kinematic modelling of high spectral resolution spectra. The CO first overtone emission has been observed toward G033.3891 with VLT/CRIRES \citep{Ilee2013}, providing a relatively high spectral resolution view of the bandhead features ($R\sim30$,0000). In that study, the $v=2$--0 bandhead was well reproduced by a model of a thin Keplerian disc at an inclination of $40\degr$ with an inner radius of 2.1\,au\footnote{Due to the limited wavelength range able to be fit with the model, the uncertainties on these parameters were in some cases undefined.  See \citet{Ilee2013}, their Section 3.3 for further details.} (Figure~\ref{G033-CRIRES}).

We note that the stellar parameters describing G033.3891 used in the previous work to fit the CRIRES spectra were taken from a preliminary collation of information in the RMS survey \citep{Lumsden2013}. In the meantime, the improved estimate of the bolometric luminosity of the source has increased from $1.0\times10^{4}$\,L$_{\odot}$ to $1.3\times10^{4}$\,L$_{\odot}$. The expected change in stellar mass resulting from this difference is of the order 0.5\,M$_{\odot}$ \citep[note the estimated mass of 12.3\,M$_{\odot}$;][]{Ilee2013}, which would correspond to changes in orbital velocity of less than 1.5\,km\,s$^{-1}$ beyond 2\,au. Since this is far less than the spectral resolution of the observations ($\Delta v \sim 10$\,km\,s$^{-1}$), this slight change does not affect the final results of the inner radius of the gaseous disc. 

The inner size of the gaseous disc traced with CO bandhead emission when derived independently using spatial interferometric observations (2.8~au) is in alignment with the size derived when using kinematic models applied to high spectral resolution spectra (2.1~au). The inner radius derived from both methods is independent of the inclination angle, as it represents an absolute distance from the central star rather than a projected value. While different orientations can influence detailed interpretations, the primary focus of this study is the consistency between the inner radius of the CO obtained from kinematic modelling and that derived from geometric modelling. This key result is robust and independent of the assumed inclination.

The outer radius of the CO gaseous region could not be accurately constrained by either kinematic or geometric modelling. The kinematic modelling set the outer radius as far out as 3200~au. This value is a result of extrapolation using the temperature exponent, which in turn is poorly defined when fitting only the 2-1 bandhead. Combined with the inability of the models to set an uncertainty, we conclude that there is no real evidence that the CO gaseous ring extends that far out. The geometric models applied to GRAVITY observations suffered from a similar uncertainty regarding the fit of the outer radius. 

Fitting the rest of the bandheads is crucial for constraining the outer radius of the disc and confirming its true extent, especially since the first overtone emission is known to be optically thick \citep[e.g.,][]{Gra_koutoulaki_2021}. The GRAVITY spectrum could aid in this task, but the combination of it being heavily affected by the atmosphere and being of a lower resolution compared to the CRIRES spectrum makes it very challenging. Probing the fundamental CO emission in the mid-infrared wavelengths (e.g., with MATISSE) in the future, is therefore important in constraining the outer radius of the disc. 

Our study highlights the powerful impact of combining spatial and kinematic data. This approach helps us pinpoint with great accuracy where the CO bandhead emission originates from, revealing the closest region of gas accretion around this MYSO, well within the sublimation radius. Notably, our analysis of G033.3891 marks the second instance of such investigation in MYSOs, contributing significantly to our understanding of how these massive young stars gather mass.



\subsection{Spectro-astrometry}

For the Br$\gamma$ emission, for which clear variations along the differential phases are observed, it is possible to obtain an astrometric solution. An astrometric solution allows the precise measurement of both the position and the motion of the region emitting the Br$\gamma$ line, informing us about its spatial distribution and dynamics. To constrain the displacement of the photocentre for each velocity channel of the Br$\gamma$ emission we used the calibrated differential phases following the equations described in \cite{Caratti2016}. To ensure that the photocentre offsets correspond to the line-emitting region, the continuum contributions are subtracted from both the visibilities and the differential phases. The displacement of the photocentre of the emission at any given wavelength is then computed as:

\begin{equation}
\delta = -\Delta \Phi \frac{\lambda}{2 \pi B}
\end{equation}

\hspace{-0.5cm}where $\Delta$$\Phi$ is the continuum subtracted differential phase and $B$ is the length of the baseline \citep{Lachaume2003,LeBouquin2009}. The displacements are calculated for all available baselines and for each spectral channel over the line profile where line-to-continuum ratio is greater than 10\%. The velocities are calculated with respect to the local standard of rest, which is 26.85 km/s. 

\begin{figure}
    \centering
    \includegraphics[width=0.5\textwidth]{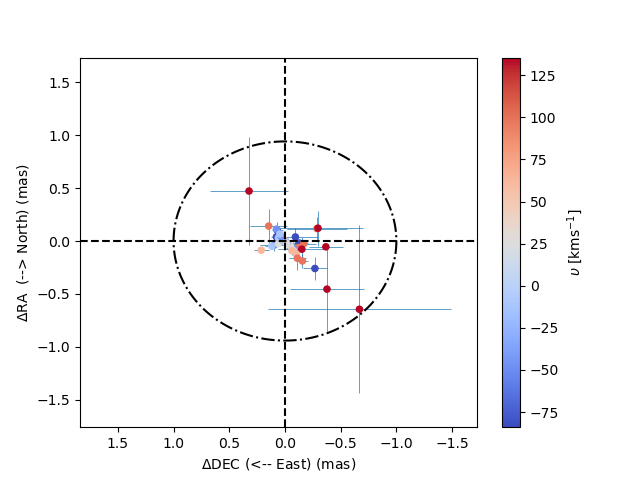}
    \caption{Br$\gamma$ photo-centre shift displacements for different velocity channels calculated from the differential phases after correcting for the continuum contribution. The black contour corresponds to the inner ring size of the continuum emission.}
    \label{G033-astrometry}
\end{figure}

Figure~\ref{G033-astrometry} shows that the displacement of the photocentre of the blue and red shift velocities generally follows a distribution where the blue-shifted components are mostly located towards the North-East, and the red-shifted components are mostly located towards South-West. The exception to this trend is three red-shifted data points towards the east and one blue-shifted data point towards the west. We note that the observed blend of red and blue-shifted velocity components could be due to the spectral resolution. The reported velocities range between -75 km/s and 125 km/s. If the observed velocity discrepancies are real, they add an extra complexity to interpreting the physical origin of the emission, which altogether seems to follow the asymmetric direction observed in the continuum emission. The PAs of the continuum asymmetry and the offset from the photocentre of the Br$\gamma$ emission are found to be 70$^{\circ}$ and 41$^{\circ}$ respectively. 

The global astrometric solution follows the geometry we obtained in Sect.~\ref{geo_br}, where the brightness distribution of the continuum and the displacement of the Br$\gamma$ are both stronger towards North-East. 

The obtained astrometric solution of the Br$\gamma$ emission indicates that if attributed to a jet, it diverts from a perfect edge-on jet geometry. An inclined disc could explain the observed asymmetry of the assumed jet, as well as the decrease in the brightness of the dusty disc in the South-West at PA of 250$^{\circ}$ (i.e., an obscured inclined disc).   

In conclusion, our findings suggest that there is a mix of a low and a high velocity moving ionised gas towards the East-North, West-South orientation, following a similar orientation of the dusty disc. The lack of a systematic trend in the velocity field indicates that Br$\gamma$ emission is probably a result of both a disc/wind and a jet mechanism within the inner 0.5~mas, while the emission becomes more collimated outwards (i.e., outflow/jet origin). 

\section{Discussion}

\subsection{Br\texorpdfstring{$\gamma$}{} emission}

The origin of the Br$\gamma$ line in (M)YSOs is a matter of long-term debate, with various proposed physical mechanisms that could give rise to this emission. These mechanisms include the accretion of matter onto the star, outflowing material originating from a disc wind, an extended wind or outflow, or a jet \citep{Eisner2007,Garcia2015,Stecklum2012,Koumpia2021}.

Our geometric modelling combined with the astrometry obtained from the HR GRAVITY/VLTI observations of G033.3891 suggests that the main component of the Br$\gamma$ emission has a bipolar origin. The mix of low/high-velocity components and blue/red-shifted emission at similar distances from the central star indicates that the mechanism responsible for the Br$\gamma$ emission is rather a combination of disc/wind and a jet. We report velocities that exceed $\sim$100~km/s, while noting that jets are expected to reach typical velocities as high as several hundred km/s \citep[e.g.,][]{Torrelles2011}. 


The general observation of increasing velocities with increasing distance from the central source can exclude a pure accreting origin of the emission (e.g., magnetospheric accretion). The estimated size of the Br$\gamma$ emission is also larger by a factor between 4-40 compared to the Alfvén radius, assuming an approximation of the magnetospheric radius to be in the range of 2-20~$R_{*}$ \citep[see,][]{Hartmann2016, Garcia2020, Zhu2024}. Both the size, the velocity information and the geometry of the Br$\gamma$ emission cannot support that magnetospheric accretion can be its sole responsible underlying mechanism. 

The visibility and differential phases observed at high velocities suggest that the outflowing material spans up to $\sim$3~au, therefore, the emission is quite compact. We note that for a jet one would expect material that is more extended reaching a few tens of au \citep{Caratti2016}, which is within the range of scales that GRAVITY can trace. Initial jet opening angles of $\sim$ 30 degrees are seen in collimated material around low-mass YSOs. The astrometry of Br$\gamma$ around G033.3891 indicates a blend of both collimated and more dispersed emissions, further supporting a joint origin of jet and a disc/wind as the underlying cause of the Br$\gamma$ emission. 

\subsection{Molecular gas emission}

This is the third case of an MYSO, showing spatially resolved CO bandhead emission \cite{Caratti2020, Koumpia2023} and the second ever study to combine geometric and kinematic modelling to trace its exact origin. 

Geometric models reveal that the CO emission may stem from a ring tighter to the central protostar compared to the dust. The CO is somehow protected from undergoing photo-dissociation, a process that would typically occur due to the exposure to intense ultraviolet (UV) photons by the central stellar object at such proximity. This is probably due to the CO gas emitting disc, self-shielding itself. 

The CO ring that we found has an inner radius of around 2.8~au which within uncertainties, matches our results from the kinematic modelling of the CRIRES spectra (R$\sim$30,0000) towards the source \cite[2.1 au, see also,][]{Ilee2013}. The kinematic fit assumed a circumstellar disc in Keplerian rotation and found an inclination of 40$^{\circ}$ for G033.3891. This could cause the shadowing we observed in the dusty disc via the interferometric fitting. To achieve more robust conclusions, it would be necessary to include the inclination and position angle of the disc in the fitting process. This could also refine the size and error estimates, but a fuller uv-plane is necessary to aid these efforts. Constraining the outer radius of the gaseous disc via the CO bandhead emission requires the observation of optically thin emission, better traced with higher-order CO bandheads (e.g., third-order traced by CRIRES+). 

We note that the most important finding of the current study concerns the constraint of the inner radius of the gaseous CO disc, as it proves active accretion via Keplerian motion, at a location of only a few aus away from the central MYSO.

\section{Conclusions}

This paper presents the geometric modelling of the first K-band interferometric observations of the MYSO G033.2891 (GRAVITY/VLTI) combined with kinematic modelling of its former high spectral resolution CRIRES observations of the CO bandhead emission. G033.2891 is the third MYSO known to have spatially resolved CO (together with IRAS 13481-6124 and NGC 2024 IRS 2), adding an object to the sparse sample of such observations. The present CO analysis marks only the second investigation of its kind within MYSOs, reinforcing the disc-like nature of accretion at MYSOs at the smallest scales. In addition, the differential phases around the Br$\gamma$ emission allow us to perform spectro-astrometric analysis and further investigate its physical origin.  

The main conclusions concern the sizes of the 2.2~$\mu$m, the CO, and the Br$\gamma$ emissions and are summarised as follows.  

\begin{enumerate}
  \item The 2.2~$\mu$m continuum emission arises from a ring with an inner radius consistent with the computed dust sublimation radius ($\sim$2~au).
  \item The inner radius of the CO emission as determined spatially, arises from an area significantly smaller ($\sim$2.8~au) than the dusty disc ($\sim$5~au) and, within the errors, matches the inner radius determined independently from the kinematic modelling assuming Keplerian motions ($\sim$2.1~au). 
  \item The astrometric solution of the Br$\gamma$ emission suggests a mixture of collimated and scattered emissions without apparent systematics in the velocity field. This finding suggests that both a jet and a disc/wind are jointly responsible for generating the observed Br$\gamma$ emission.
\end{enumerate}

The importance of CO bandhead emission in understanding accretion/ejection in MYSOs is increasingly acknowledged. To fully grasp its extent, combining spatial data with high-resolution spectra covering the optically thin bandheads is crucial. Including species like Na I, CO, and Br$\gamma$ can refine kinematic information, by constraining the velocity gradient of the neutral, ionised and molecular content of the star/disc interface. As individual MYSO studies multiply, emphasizing larger, statistically significant sample sizes becomes apparent (e.g., CRIRES+, GRAVITY+). Spectro-astrometry is a powerful tool to constrain the physical origin of the emission lines, urging future studies to leverage this technique for a deeper understanding of the accretion/ejection processes around MYSOs.

\begin{acknowledgements}

We thank the anonymous referee for their detailed feedback, which greatly improved the quality of this manuscript. We also acknowledge Antoine M\'erand for his assistance with PMOIRED.
JDI acknowledges support from an STFC Ernest Rutherford Fellowship (ST/W004119/1) and a University Academic Fellowship from the University of Leeds. 
Based on observations collected at the European Southern Observatory under ESO programme 0105.C-0686(D) (GRAVITY). 

\end{acknowledgements}





\bibliographystyle{aa} 
\bibliography{example}




\end{document}